\documentstyle[12pt]{article}
\newcommand{\beq}{\begin{equation}}
\newcommand{\eeq}{\end{equation}}
\newcommand{\beqa}{\begin{eqnarray}}
\newcommand{\eeqa}{\end{eqnarray}}
\def\nue{{\nu_e}}

\def\numu{{\nu_{\mu}}}

\def\nutau{{\nu_{\tau}}}

\newcommand{\dm}{\mbox{$\Delta{m}^{2}$~}}
\newcommand{\st}{\mbox{$\sin^{2}2\theta$~}}
\begin{document}
\begin{center}
{\large{ \bf {Is 
neutrino decay really ruled out as a solution to the atmospheric
neutrino problem from Super-Kamiokande data?}}} \\
\vskip 10pt
{\it  Sandhya Choubey$^{a}$ and Srubabati Goswami$^{b}$ \\
$^{a}$ Saha Institute of Nuclear Physics,\\1/AF, Bidhannagar,
Calcutta 700064, INDIA.\\
$^{b}$ Physical Research Laboratory, Ahmedabad 380009, INDIA}
\vskip 30pt

{\bf{Abstract}}
\end{center}
In this paper we do a detailed $\chi^2$-analysis of the 
848 days of Super-Kamiokande(SK) 
atmospheric neutrino data
under the assumptions of $\nu_\mu - \nu_\tau$ 
oscillation and neutrino 
decay. For the latter we take the most general case of neutrinos 
with non-zero mixing and consider the possibilities of 
the unstable component in $\nu_\mu$ decaying 
to a state with which it mixes (scenario (a)) and to a sterile state 
with which it does not mix (scenario (b)). In the first case 
$\Delta m^2$ (mass squared difference between the two mass states 
which mix) has 
to be $>$ 0.1 $eV^2$ from constraints on $K$ decays while for the 
second case $\Delta m^2$ can be unconstrained. 
For case (a) \dm does not enter 
the $\chi^2$-analysis while in case (b) it enters the 
$\chi^2$-analysis as an independent parameter. 
In scenario (a) there is \dm averaged oscillation in addition to decay
and this gets ruled out at 100.0\% C.L. by the latest SK data. 
Scenario (b) on the other hand gives a reasonably good 
fit to the data for \dm  $\sim 0.001 ~eV^2$.

{\it keywords:} neutrinos; mixing; oscillations; decay

\vskip 60pt
--------------------\\
$^{a}$ e-mail: sandhya@tnp.saha.ernet.in \\
       phone : 011-91-33-3375345\\
       fax : 011-91-33-3374637\\
$^{b}$ e-mail: sruba@prl.ernet.in

\newpage


\section{Introduction}

The recent data of Super-Kamiokande (SK) \cite{sk} have given a new 
impetus to the atmospheric neutrino problem and a possible                      
interpretation in terms of neutrino  oscillation. 
Moreover the high statistics of SK makes it possible to
study the zenith-angle dependence of the neutrino flux from which
one can conclude that the $\nu_\mu$'s show signs of oscillation but the 
$\nu_e$ events are consistent with the no-oscillation hypothesis. 
Independently the results from the reactor experiment CHOOZ disfavours the 
$\nu_\mu - \nu_e$ oscillation hypothesis \cite{chooz}. On the other hand  
large angle $\nu_\mu-\nu_\tau$ or $\nu_\mu-\nu_s$ 
($\nu_s$ being a sterile neutrino) solution continues to give a 
good fit to the data. Nevertheless effort has been on to try out 
other possibilities to explain the anomaly observed in SK and one among these 
is neutrino decay \cite{lipari,pak}. 
In \cite{lipari} it was shown that neutrino decay gives a poor fit to the
data. However they considered neutrinos with zero mixing. 
Barger {\it et al.} considered the situation 
of neutrino decay in the general case of neutrinos with non-zero mixing angle
\cite{pak}. 
They showed that the neutrino decay fits the $L/E$ distribution of the SK
data well. The $\Delta m^2$ taken by them was $>$  0.1 $eV^2$ so that the 
$\Delta m^2$ dependent term averages out. 
As pointed out in \cite{pak} such a constraint on \dm is valid when the 
unstable  state decays into some other state with which it mixes. If however
the unstable state decays into a sterile state with which it does not mix
then there is no reason to assume \dm $>0.1~eV^2$. 

In this paper we present our results of two-flavour $\nu_\mu - \nu_\tau$ 
oscillation and neutrino decay solutions to the atmospheric neutrino
problem by  
doing $\chi^2$-fit to the 848 days of 
sub-GeV and multi-GeV Super-Kamiokande data \cite{kate}.
We have also presented the results of $\chi^2$-fit to the 
535 days SK data and have compared it with the results for the 
new data. 
For the neutrino decay analysis we take the most general case of
neutrinos with non-zero mixing and consider two pictures 
\begin{itemize}
\item \dm $>$ 0.1 $eV^2$ (scenario (a))
\item \dm unconstrained  (scenario (b))
\end{itemize}
We also explicitly 
demonstrate the behavior of the up-down asymmetry parameters 
\cite{fl,yasuda} in both scenarios.
 
Our analysis shows that scenario (a) 
is ruled out at 100\%(99.99\%) C.L. by the 848(535) days of SK data. 
However if we remove the constraint on $\Delta m^2$ 
and consider the possibility of decay into a sterile 
state
then one can get an acceptable fit for $\Delta m^2$ $\sim 0.001 eV^2$ and
\st large.

The plan of the paper is as follows. In section 2 we present our results
for two-generation $\nu_\mu-\nu_\tau$ oscillation analysis. 
In section 3.1 we present our results for the neutrino decay
solution constraining \dm to be $> 0.1 eV^2$. 
In section 3.2 we do a three parameter 
$\chi^2$ analysis by removing the constraint on
$\Delta m^2$.
In section 4 we perform a comparative study of the three cases and
indicate how one can distinguish experimentally between the scenario
(b) and the $\nu_\mu - \nu_\tau$ oscillation case though both give almost
identical zenith-angle distribution. 

\section{$\nu_\mu - \nu_\tau$ oscillation}

In the two-flavour picture the probability that an initial $\nu_{l}$ 
of energy
$E$ remains a  $\nu_{\l}$ after traveling a distance $L$
in vacuum is
\begin{equation}
P_{\nu_{l}\nu_{l}} =1- \sin{^2}2\theta \sin{^{2}}(\pi L/\lambda_{osc})
\label{p2nu}
\end{equation}
where $\theta$ is the mixing 
angle between the two neutrino states in vacuum and 
$\lambda_{osc}$ is the oscillation wavelength defined as,
\begin{equation}
\lambda_{osc} = (2.5~ km)~ \frac{E}{GeV} \frac{eV^2}{\Delta m^2}
\label{lo}
\end{equation}
where $\Delta m^2$ denotes the mass squared difference between the two
mass eigenstates.  
The expected number of $l$ (e or $\mu$) like 1 ring events recorded 
in the detector in presence of oscillations is given by
\begin{eqnarray}
N_l
& = & n_T
\int^{\infty}_{0} dE
\int^{(E_l)_{\rm max}}_{(E_l)_{\rm min}} dE_l
\int_{-1}^{+1} d\cos \psi
\int_{-1}^{+1} d\cos \xi\
{1 \over 2\pi}
\int_{0}^{2\pi} d\phi
\nonumber\\
&\times&
{d^2F_l (E,\xi) \over dE~d\cos\xi}
\cdot{ d^2\sigma_l (E,E_l,\cos\psi) \over dE_l~d\cos\psi } 
\epsilon(E_l)
\cdot
{\ }P_{\nu_l \nu_l} (E, \xi).
\label{rate}
\end{eqnarray}
$n_T$ denotes the number of target nucleons, $E$ is the neutrino energy, 
$E_l$ is the energy of the final charged lepton, $\psi$ is the angle 
between the incoming neutrino $\nu_l$ and the scattered lepton $l$, 
$\xi$ is the zenith angle of the neutrino and $\phi$ is the azimuthal angle 
corresponding to the incident neutrino direction (the azimuthal angle 
relative to the $\psi$ has been integrated out). The zenith angle of the 
charged lepton is then given by 
\begin{equation}
\cos \Theta = \cos \xi \cos \psi + \sin \xi \cos \phi \sin \psi
\label{zenith}
\end{equation}
$d^2F_l /dE d\cos\xi$ is the differential flux of atmospheric
neutrinos of type $\nu_l$, 
$d^2\sigma_l/dE_l d\cos\psi$ is the differential cross section
for $\nu_l N \rightarrow l X$ scattering and $\epsilon(E_l)$ 
is the detection efficiency for the 1 ring events in the detector. 
The efficiencies that were available to us are not the detection 
efficiencies 
of the charged leptons but some function which we call $\epsilon(E)$
defined as \cite{private}
\begin{equation}
\epsilon(E) = \frac{ \int{ \frac{d\sigma}{dE_l} \epsilon(E_l)
dE_l}}
{\int{ \frac{d\sigma}{dE_l} dE_l}}
\end{equation} 
$P_{\nu_l \nu_l}$ is the survival probability of a 
neutrino flavour $l$ after traveling a distance $L$ given by, 
\begin{equation}
L= \sqrt{(R_e+h)^2 - {R_e}^2 \sin^2 \xi} - R_e \cos \xi
\end{equation}
$R_e$ being the radius of the earth and h is the height of the atmosphere
where the neutrinos are produced.
We use the atmospheric neutrino fluxes from \cite{honda}. 
For the sub-GeV events the dominant process is the charged current 
quasi-elastic scattering from free or bound nucleons. 
We use the cross-sections given in \cite{gaisser}.   
The events in multi-GeV range have contributions coming from 
quasi-elastic scattering, single pion production and multi pion production and 
we have used the cross-sections given in \cite{lipary}. For the multi-GeV 
events we assume that the lepton direction $\Theta$ is the same as the 
incoming neutrino direction $\xi$. But actually they are slightly 
different. We simulate this difference in the zenith angles by 
smearing the angular distribution of the number of events with a 
Gaussian distribution having a one sigma width of $15^{\rm o}$ for 
$\mu$ type events and $25^{\rm o}$ for the e type events \cite{G_G}.
For the sub-GeV 
events, difference in direction between the charged lepton and the neutrinos 
are exactly taken care of according to eq. (\ref{rate}) and (\ref{zenith}). 

To reduce the uncertainty in the absolute flux values 
the atmospheric neutrino measurements are usually presented in terms of 
the double ratio 
\begin{equation}
R = {\frac{(\nu_\mu + \overline{\nu}_{\mu})/
(\nu_e + \overline{\nu}_e)_{\rm obsvd}}
{(\nu_\mu + \overline{\nu}_{\mu})/(\nu_e + \overline{\nu}_e)_{\rm MC}}}
\label{ratm}
\end{equation}
where {MC} denotes the Monte-Carlo simulated ratio.
Different calculations agree to within better than 5\% on the magnitude
of this  quantity.
We use a similar quantity R, where
\begin{equation}
R \equiv \frac{(N_{\mu}/N_e)|_{osc}}{(N_{\mu}/N_e)|_{no-osc}}.
\end{equation}
The quantities $N_{e,\mu}$ are the numbers of $e$-like and
$\mu$-like events, as per eq.(\ref{rate}).
The numerator denotes numbers obtained from eq.(\ref{rate}), while the
denominator the numbers expected with the survival probability 
as 1.

At the detector, the neutrino flux come from all directions.
Thus, the total path length between the
production point in the atmosphere and the detector varies 
from about 10 km to 13,000 km depending on the zenith angle.
Neutrinos with zenith angle less than $90^{\rm o}$ ({\it
`downward neutrinos'}) travel a distance of $\sim$ 10 -- 100 km from their
production point in the atmosphere to the detector while the neutrinos
with larger zenith angles ({\it `upward neutrinos'}) cross a distance of
up to $\sim$ 13,000 km to reach the detector. 
Apart from altering the
flavour-content of the atmospheric neutrino flux, oscillations could
lead to the following effect: if the oscillation length is
much longer than the height of the atmosphere but smaller than the
diameter of the earth, only upward neutrinos coming from the opposite
side of the earth will have significant oscillations. These would show
up as an up-down asymmetry in the event distribution.
SK has enough statistics to study these up-down flux asymmetry. 
They divide the $(-1,+1)$ interval in  $\cos\Theta$ in five equal bins:
$(-1.0,-0.6)$, $(-0.6,-0.2)$, $(-0.2,+0.2)$, $(+0.2,+0.6)$, $(+0.6,+1.0)$
and give the number of events in each bin. The first two bins
correspond to the upward neutrinos and the last two bins correspond to
the downward neutrinos. 
To probe the up-down flux asymmetries we use the parameter $Y$ defined 
in \cite{yasuda},
\begin{equation}
Y_{l} \equiv {(N_{l}^{-0.2}/N_{l}^{+0.2})|_{osc}
\over (N_{l}^{-0.2}/N_{l}^{+0.2})|_{no-osc}}.
\end{equation}
Here $N_{l}^{-0.2}$ denotes the number of $l$-type events
produced in
the detector with zenith angle $\cos \Theta < -0.2$, {\it i.e.} the upward
neutrino events while
$N_{l}^{+0.2}$ denotes the number of $l$-type events for $\cos \Theta >
0.2$ {\it i.e.} events coming from downward neutrinos. 
The central bin has contributions from both upward and downward
neutrinos and is not useful for studying the 
up-down asymmetry. 

We minimize the $\chi^2$ function defined as  \cite{yasuda}
\begin{equation}
\chi^2 = \sum_{i} \left[\left({R}^{exp} - R^{th} \over \delta
R^{exp} \right)^2 
+ \left({Y^{exp}_{\mu} - Y^{th}_{\mu} \over \delta Y^{exp}_{\mu}}\right)^2
+ \left({Y^{exp}_{e} - Y^{th}_{e} \over \delta Y^{exp}_{e}}\right)^2
\right],
\label{chi}
\end{equation}
where the sum is over the sub-GeV and multi-GeV cases.
The experimentally observed rates
are denoted by the superscript "exp"
and the theoretical predictions for the quantities are labeled by "th".
$\Delta R^{exp}$ is the
error in $R$ obtained by combining the statistical and
systematic errors in quadrature.
$\Delta Y^{exp}$ corresponds to the error in $Y$. For this we take only
the statistical errors since these are much larger compared to the
systematic errors. 
We include both the $e$-like and the $\mu$-like up-down asymmetries in the
fit so that we have 4 degrees of freedom (6 experimental data - 2
parameters) for the oscillation analysis in the two parameters 
$\Delta m^2$ and $\sin^2 2\theta$.

The use of these type of ratios for the $\chi^2$ analysis test has been 
questioned in \cite{lisi1} because the error distribution of these ratios is
non-Gaussian in nature. The alternative is to use the absolute number of
e or $\mu$ type events taking into account the errors and their
correlations properly \cite{G_G,lisi2}. However as has been shown in 
\cite{yasuda} the use of the $R$'s and $Y$'s as defined above is 
justified within the 3$\sigma$ region around the best-fit point for 
a high statistics experiment like SK and provides an alternative way
of doing the $\chi^2$-analysis.
A comparison of the results of \cite{yasuda} with those obtained in 
\cite{G_G,lisi2} shows that the best-fit points and the allowed regions
obtained do not differ significantly in the two approaches of data
fitting. 
The advantage of using the ratios is that they are relatively 
insensitive to the uncertainties in the neutrino fluxes and
cross-sections as the overall normalization factor gets canceled out 
in the ratio. We have included the $Y_e$ in our analysis because
to 
justify the $\nu_\mu - \nu_\tau$ oscillation scenario, it is necessary to
check that $\chi^2$ including the data on electron events  gives a low value
and hence it is the standard practice to
include these in the $\chi^2$-analysis \cite{yasuda,G_G,lisi2}.
The data that we have used are shown in Table 1 which corresponds to the
848 days \cite{kate} and the 535 days \cite{data} of data. 

\begin{description}
\item{Table 1:} The SK data used in this analysis.
\end{description}
\[
\begin{array}{|c|c|c|c|c|} \hline
{} & \multicolumn {2}{c|} {\rm 848 ~ days ~ data} & 
\multicolumn {2}{c|} {\rm 535 ~ days ~ data} \\ \cline {2-5}
{\rm Quantity} & {\rm Sub-GeV} & {\rm Multi-GeV} 
& {\rm Sub-GeV} & {\rm Multi-GeV} \\ \hline 
{R^{exp}} & {0.69} & {0.68} & {0.63} & {0.65} \\ \hline
{\Delta R^{exp}} & {0.05} & {0.09} & {0.06} & {0.09} \\ \hline
{Y^{exp}_\mu} & {0.74} & {0.53} & {0.76} & {0.55} \\ \hline
{\Delta{Y^{exp}_\mu}} & {0.04} & {0.05} & {0.05} & {0.06} \\ \hline
{Y^{exp}_e} & {1.03} & {0.95} & {1.14} & {0.91} \\ \hline
{\Delta {Y^{exp}_e}} & {0.06} & {0.11} & {0.08} & {0.13} \\ \hline
\end{array}
\]

For the 2 flavour $\numu-\nutau$ oscillation 
the $\chi^2_{min}$ that we get is 1.21 with the best-fit values as
\dm = 0.003 $eV^2$ and \st = 1.0. 
This provides a good fit to the data being allowed at 87.64\% C.L. 
If we use the 535 days data then 
the $\chi^2_{min}$ that we get is 4.25 with the best-fit values as
\dm = 0.005 $eV^2$ and \st = 1.0, the g.o.f being 37.32\%. 
Thus the fit becomes much better with the 848 days data with 
no significant change in the best-fit values. 
Though we have used a different procedure of data fitting, our results
agree well with that obtained by the SK 
collaboration\footnote{The best-fit values that the 
SK collaboration has got for the
848 days data are \cite{kate}
\dm = 0.003 eV$^2$, \st = 0.995 and $\chi^2/d.o.f$ = 55.4/67.
This corresponds to a g.o.f of 84.33\%.}. 
In fig. 1 we show the 90\% C.L.  
($\chi^2 \leq \chi^2_{min} + 4.61$) and the 99\% C.L. 
($\chi^2 \leq \chi^2_{min} + 9.21$) allowed region  
in the (\dm, \st) plane for the $\nu_\mu - \nu_\tau$ oscillation hypothesis 
using the latest SK data.

\section{Neutrino decay}

The neutrino decay hypothesis assumes that there is an unstable
component in  
$\numu$ (say $\nu_2$) which decays into one of the lighter 
states (say $\nu_3$).
Experimental considerations constrain $\nue$ to decouple from 
$\nu_2$ and it's decay partners, so that 

\begin{equation}
\nue \approx \nu_1
\end{equation}
\begin{equation}
\nu_\mu \approx \nu_2 \cos\theta + \nu_3 \sin\theta
\label{nm}
\end{equation}
From (\ref{nm}) the survival probability of the $\numu$ of energy E, 
with an unstable component $\nu_2$,
after traveling a distance $L$ is given by,
\begin{eqnarray}
P_{\numu\numu} &=& \sin^4\theta + \cos^4\theta \exp(-4 \pi L/\lambda_d)\nonumber\\
&& {}+ 2\sin^2\theta \cos^2\theta \exp(-2 \pi L/\lambda_d) \cos(2 \pi L/\lambda_{osc})\,,
\label{pmumudo}
\end{eqnarray}
where $\lambda_d$ is the decay length
(analogous to the oscillation wavelength given by eq. (\ref{lo})) defined as,
\begin{equation}
\lambda_d = 2.5 km \frac{E}{GeV} \frac{eV^2}{\alpha}
\label{ld}
\end{equation}
and $\alpha = m_2/\tau_0$, $m_2$ being the mass of the state $\nu_2$ and 
$\tau_0$ the decay lifetime. The $\lambda_{osc}$
appearing in eq. (\ref{pmumudo}) is the wavelength of oscillations 
as defined in eq. (\ref{lo}) with \dm = $m_2^2 - m_3^2$.

\subsection{$\Delta m^2 > 0.1 eV^2$}

If the unstable component in the $\nu_\mu$ state decays to some other state
with which it mixes then bounds from $K$ decays imply 
$\Delta m^2 > 0.1 eV^2$ \cite{pak2}. 
In this case the $\cos(2 \pi L/\lambda_{osc})$ term
averages to zero and the probability becomes
\begin{equation}
P_{\numu\numu} = \sin^4\theta + \cos^4\theta \exp(-4 \pi L/\lambda_d) \,.
\label{pmumu}
\end{equation}

In figs. 2 and 3 we show the variation of $R$ and $Y$ with $\alpha$ for
various values of $\sin^2 \theta$ for the sub-GeV and multi-GeV cases. 
For higher values of $\alpha$, the decay length $\lambda_d$ 
given  by eq. (\ref{ld}) is low and the exponential term in the survival
probability is less implying that more number of neutrinos decay 
and hence  $R$ is
low.  As $\alpha$ decreases the decay length increases and the number 
of decaying neutrinos decreases, increasing $R$. 
For very low values of $\alpha$ 
the exponential term goes to 1, the neutrinos do not get the time to 
decay so that the 
probability becomes $1-\frac{1}{2}\sin^2 2\theta$ and remains constant 
thereafter for all lower values of $\alpha$. 
This is to be contrasted with the $\nu_\mu-\nu_\tau$ oscillation case 
where in the no oscillation limit the $\sin^2 (\pi L/\lambda_{osc})$ term
$\rightarrow$ 0 and the survival probability $\rightarrow$ 1.  
For multi-GeV neutrinos since the energy is higher the
$\lambda_d$ is higher and the no decay limit is reached for a 
larger value of $\alpha$ as compared to
the sub-GeV case. This explains why the multi-GeV curves become flatter 
at a higher $\alpha$.
The behavior of the up-down asymmetry parameter is also completely
different from the only oscillation case \cite{yasudafig}. 
In particular the plateau obtained for a range of
\dm which was considered as a characteristic prediction for
up-down asymmetries is missing here.
For the decay case even for $\alpha$ as high as 0.001 $eV^2$,
the decay length $\lambda_d = 2500~(E/GeV)~km$ so that the 
exponential term is 1,  
there is almost no decay for the downward neutrinos 
and the survival probability is $P = 1 - \frac{1}{2}\sin^2
2\theta$ while the upward going neutrinos have some decay and so $Y$ is 
less than 1. As $\alpha$ decreases, the $\lambda_d$ increases, and
the fraction of upward going neutrinos decaying decreases and this increases 
$Y$. For very small values of $\alpha$ even the upward neutrinos do not
decay and $Y \rightarrow$ 1 being independent of $\theta$. 
 
We also perform a $\chi^2$ analysis of the data calculating the 
"th" quantities in (\ref{chi}) for this scenario. 
The best-fit values that we get are $\alpha = 0.33 \times 10^{-4}$ in
$eV^2$ 
and $\sin^2\theta = 0.03$ with a $\chi^2_{\rm{min}}$ of 49.16.
For 4 degrees of freedom this solution is ruled out at 100\% C.L. 
The best-fit values for the 535 days of data 
that we get are $\alpha = 0.28 \times 10^{-4}$ in
$eV^2$ 
and $\sin^2\theta = 0.08$ with a $\chi^2_{\rm{min}}$ of 31.71.
For 4 degrees of freedom this solution is ruled out at 99.99\% C.L. 
\cite{lisidecay}.
Thus the fit becomes worse with the 848 days data as compared to the
535 days data. 
We have marked the $R$ and $Y$ corresponding to the 
best-fit value of the parameters $\alpha$ and $\sin^2\theta$ in figs. 
2 and 3. It can be seen that the best-fit value of R for the sub-GeV 
neutrinos is just below and that for the multi-GeV neutrinos is just above 
the $\pm 1\sigma$ allowed band of the SK 848 days of data. The up-down 
asymmetry parameter $Y$ is quite low for the sub-GeV neutrinos and 
extremely high for the multi-GeV neutrinos as compared to that allowed 
by the data. The fig. 2 shows that for the sub-GeV neutrinos the data 
demands a lower value of $\alpha$ while from fig. 3 we see that the multi-GeV 
neutrinos need a much higher $\alpha$ to explain the SK data.
It is 
not possible to get an $\alpha$ that can satisfy both the sub-GeV and the
multi-GeV SK data, particularly it's zenith angle distribution.   
In this 
scenario, decay for the sub-GeV upward neutrinos is more than that for the 
multi-GeV upward neutrinos (downward neutrinos do not decay much) and as a 
result $Y$ for sub-GeV is lower than the $Y$ for multi-GeV, a fact not 
supported by the data. 
Since the 848 days data needs even lesser depletion of the sub-GeV flux 
as compared to the multi-GeV flux, 
the fit gets worse.  


\subsection{$\Delta m^2$ unconstrained}
In this section we present the results of our $\chi^2$-analysis removing
the constraint on $\Delta m^2$. 
This case corresponds to the unstable neutrino state decaying to some sterile
state with which it does not mix \cite{pak}.
The probability will be still given by eq. (\ref{pmumudo}).

In fig. 4 and 5 we plot the R vs. \dm and Y vs. \dm for the sub-GeV and
multi-GeV data for $\alpha$ = 0.3 $\times 10^{-5} eV^2$ (which is the best-fit 
value we get for the 848 days data) and  
compare with the curve obtained for the best-fit 
value of $\sin^2 \theta$ (=0.5) 
for the only oscillation case (solid line). 
For the best-fit value of $\alpha$ that we get, the downward neutrinos do
not have time to decay while the upward neutrinos undergo very little decay. 
Thus the curves are very similar in nature to the only oscillation
curves.
In the sub-GeV case (fig. 4), for 
high values of \dm around 0.1 $eV^2$ both upward and downward
neutrinos undergo \dm independent average oscillations and 
R stays more or less constant with \dm. 
For the upward going neutrinos in addition to average oscillation there
is little amount of decay as well and hence Y $\sim$ $N_{up}/N_{down}$
is $\stackrel{<}{\sim}$ 1. 
As \dm decreases to about 0.05 $eV^2$  the oscillation wavelength increases 
--  for upward neutrinos it is still average oscillation but for the downward 
neutrinos, 
the $\cos (2\pi L/\lambda_{osc})$ term becomes negative which corresponds to 
maximum oscillation effect
and the survival probability 
of these neutrinos decreases, and hence R decreases;
while the upward neutrinos continue to decay and oscillate at the same rate   
and $Y$ becomes greater than 1. 
As \dm decreases further, the downward neutrino oscillation wavelength becomes
greater than the distance traversed and they are converted 
less and less and thus R increases and Y decreases.
Below \dm = 0.001 $eV^2$ the downward neutrinos stop oscillating completely 
while for the upward neutrinos the $\cos (2\pi L/ \lambda_{osc})$ 
term goes to 1, and R and Y no longer 
vary with \dm.

For the multi-GeV case (fig. 5) the oscillation wavelength is more than
the sub-GeV case and for \dm around 0.1 $eV^2$ 
the $\cos (2 \pi L/\lambda_{osc})$ term stays close to 1 for the downward neutrinos; while 
the upward neutrinos undergo average oscillations and slight decay and
Y is less than 1. As \dm decreases the
downward neutrinos oscillate even less and 
the upward neutrinos also start departing from average
oscillations and hence $R$ increases and $Y$ decreases. 
Around 0.01 $eV^2$ the downward neutrinos stop oscillation while for 
upward neutrinos the oscillation effect is maximum ($\lambda \sim L/2$) and 
the $cos (2\pi L/ \lambda_{osc})$ term is $\sim$ -1
and $Y$ stays constant with \dm.  
As \dm decreases further the upward neutrino oscillation
wavelength increases and they oscillate less in number making 
both R and Y approach 1 for \dm around 0.0001 $eV^2$. 
For multi-GeV neutrinos the decay term contributes even
less as compared to the sub-GeV case.  
 
We perform a $\chi^2$ minimization in the three parameters 
\dm, \st and $\alpha$.
The best-fit values that we get are $\Delta m^2 = 0.003 eV^2$, \st = 1.0 and 
$\alpha = 0.3 \times 10^{-5} eV^2$. 
The $\chi^2$ minimum that we get is 1.11 which is an acceptable fit being 
allowed at 77.46\% C.L..
For the 535 days data 
the best-fit values that we get are $\Delta m^2 = 0.002 eV^2$, \st = 0.87 and 
$\alpha = 0.0023 eV^2$ with a $\chi^2_{\rm {min}}$ of 4.14 which is 
allowed at 24.67\% C.L..
Thus compared to the 535 days data, the fit improves immensely and 
the best-fit shifts towards the 
oscillation limit, the best-fit value of the decay constant $\alpha$ 
being much lower now. 
It is to be noted however,  
that the best-fit in this model does not come out to be
$\alpha = 0.0$, {\it viz} the only oscillation limit. 
In table 2 we give the contributions to $\chi^2$ from the $R$'s and $Y$'s
at the best-fit value of $\alpha$ and for the $\alpha$ = 0.0 case.  
 
\begin{description}
\item{Table 2:} The various contributions to the ${\chi^2}_{\rm{min}}$ at the 
best-fit value of $\alpha$ and at $\alpha$=0.0
\end{description}
\[
\begin{array}{|c|c|c|} \hline
{\rm Quantity} & {\rm \alpha = 0.3 \times 10^{-5}~eV^2} & {\rm \alpha = 0.0
~eV^2} \\ \hline 
{R^{sg}} & {0.085} & {0.021} \\ \hline
{Y^{sg}_{\mu}} & {0.011} & {0.033} \\ \hline
{R^{mg}} & {0.48} & {0.56} \\ \hline
{Y^{mg}_\mu} & {0.014} & {0.073} \\ \hline
{Y^{sg}_e} & {0.344} & {0.344} \\ \hline
{Y^{mg}_e} & {0.176} & {0.176} \\ \hline
\end{array}
\]
Thus from the contributions to $\chi^2$ 
we see that for the best- fit case 
there is improvement for the multi-GeV R and Y as 
compared to the $\alpha = 0.0$ case. The $\chi^2$ for sub-GeV Y also 
improves. 
In fig. 6 we plot the $\chi^2 - \chi^2_{\min}$ vs. $\alpha$ with
\dm and \st unconstrained.  
There are two distinct minima in this curve -- one for lower 
values and another at higher values of $\alpha$. 
The best-fit \dm in both cases is $\sim$ 0.001 $eV^2$.
In this model there are two competing processes -- oscillation and decay.
For lower values of $\alpha$ the decay length is greater than the
the oscillation wavelength and oscillation dominates. 
The decay term exp$(-\alpha L/E)$ is close to 1 and does not vary much with
the zenith distance $L$. As $\alpha$ increases the exponential term 
starts varying very sharply with $L$ and the variation is much more sharp
for the sub-GeV as compared to multi-GeV. 
This behavior is inconsistent with the data and that is why one gets a peak
in $\Delta \chi^2$ for higher $\alpha$. As $\alpha$ increases further
the exp$(-\alpha L/E)$ term goes to zero
for the upward neutrinos and there is complete decay of these
neutrinos while the downward neutrinos do not decay, the exponential
term still being 1. 
Whenever the exponential term is 0 or 1 for the upward neutrinos, 
the wrong energy dependence of this term
does not spoil the fit and  
these scenarios can give good fit to the data. 
Even though fig. 6 shows that the data allows a wide range of $\alpha$, 
we get the two distinct minima in the $\Delta \chi^2$ vs. $\alpha$ curve 
for high and low $\alpha$ values, for both the 535 days (dotted line) 
and 848 days (solid line) data. 
But while the 848 days data prefers the lower $\alpha$ limit, the 535 days 
data gives a better fit for the high $\alpha$ limit. The reason behind 
this is that for the 848 days data the R is much higher than for the 
535 days data. Hence the 848 days data prefers lower $\alpha$ and hence 
lower suppression. 

In fig. 7 we show the 90\% and 99\% C.L.  allowed parameter region in the
\dm- \st plane for
a range of values of the parameter $\alpha$. 
In fig. 8 we show the 90\% and 99\% C.L. contours in the 
$\alpha$ - \st plane fixing
$\Delta m^2$ at different  values. 
These contours are obtained from the definition 
$\chi^2  \leq {\chi^2}_{\min} + \Delta \chi^2$, with 
$\Delta \chi^2$ = 6.25 and 15.5 for the three parameter case 
for 90\% and 99\% C.L. respectively. 
The bottom left panel in fig 7 is for the best-fit value of $\alpha$.
For high $\alpha$ (the top left panel)  
no lower limit is obtained on
$\Delta m^2$, because even if \dm becomes so low so that there is no
oscillation the complete decay of upward neutrinos can explain their
depletion. 
As we decrease $\alpha$ the allowed parameter region shrinks and finally for
$\alpha=0$ we get the two parameter limit modulo the small difference in the
C.L. definitions for the two and three parameter cases. The upper right 
panel of fig. 8 corresponds to the best-fit value of \dm. 
For very low $\alpha$, even though there is 
no decay, we still have oscillations and that ensures that when \dm is large 
enough there is no lower bound on $\alpha$ as evident in the fig. 8. For 
\dm$=10^{-4} eV^2$ the neutrinos stop oscillating and hence we get a lower 
bound on $\alpha$ beyond which the depletion in the neutrino flux is not 
enough to explain the data.  

\section{Comparison and Conclusion}
In fig. 9 we show the histogram of the muon event distributions for the 
sub-GeV
and multi-GeV data under the assumptions of $\nu_\mu-\nu_\tau$ oscillation,
and the two scenarios of neutrino decay for the best-fit values of the 
parameters both for the 535 and the 848 days of data. 
From the figs. it is clearly seen that the
scenario (a) (big dotted line) ($\dm > 0.1 eV^2$) does not fit the data well
there being
too much suppression for the sub-GeV upward going neutrinos and too less
suppression for the multi-GeV upward going neutrinos.
The scenario (b) (\dm unconstrained, small dashed line), however,  
reproduces the
event distributions well. However with the 848 days data the sub-GeV events are
reproduced better as compared to the 535 days data and the quality of the
fit improves.

The neutrino decay is an interesting idea as
it can preferentially suppress the upward $\nu_\mu$ flux and can cause
some up-down asymmetry in the atmospheric neutrino data.
However the intrinsic defect in the decay term $\exp(-\alpha L/E)$ is that
one has more decay for lower energy neutrinos than for the higher energy ones.
Thus neutrino decay by itself fails to reproduce the observed
data \cite{lipari}.
If  however one considers the most general case of neutrinos with
non-zero mixing then there are three factors which control
the situation
\begin{itemize}
\item the decay constant $\alpha$ which determines the decay rate
\item the mixing angle $\theta$ which determines the proportion
of neutrinos decaying and mixing with the other flavour
\item the \dm which determines if there are oscillations as well
\end{itemize}

If the heavier state decays to a state with which it mixes then
\dm has to be $> 0.1 eV^2$  because of bounds coming from $K$ decays 
\cite{pak2}. 
The best-fit value of $\alpha$ that one gets is $0.33\times 10^{-4} eV^2$ 
with the latest SK data. At this value of $\alpha$ the  
$e^{-\alpha L/E}$ term tends to 1  
for the downward going neutrinos signifying
that they 
do not decay much. The survival probability  
goes to ($1 - \frac{1}{2} \sin^2 2 \theta $) which is
just the average oscillation probability.
In order to suppress this average oscillation  
the best-fit value of $\sin^2 \theta$ comes out to be small in this picture.
For the upward going neutrinos, in scenario (a),
there  will be both decay and average
oscillations. If one had only average oscillation then the probability would
have stayed constant for a fixed value of the mixing angle $\theta$. 
But because of the exponential decay term the survival probability 
drops very sharply as we go towards $\cos\Theta$=-1.0.
The drop and hence the decay is more for lower energy neutrinos. 
As a result the sub-GeV flux gets more depleted than the multi-GeV flux, 
a fact not supported by the data. In fact the  
848 days data requires the sub-GeV flux to be even less suppressed 
than the multi-GeV flux as compared to the 535 days data and the fit 
worsens with the 848 days data.  
The small mixing signifies that the
$\nu_\mu$ has a large fraction of the unstable component
$\nu_2$ (see eq. (\ref{nm})).
Hence the constant $\alpha$ comes out to be low so that
the decay rate is less  to compensate this. 
However even at the best-fit $\alpha$ of 0.33 $\times 10^{-4} eV^2$  
the survival probability in the bin with $\cos \Theta$ between -1.0 to -0.6 
comes out to be 0.15 for E=1 GeV, much lower than the value of
$\sim$ 0.5 as required by the data. 
Thus scenario (a) fails to explain the upward going neutrino data properly
because of two main reasons
\begin{itemize}
\item{$\theta$ is low in order to suppress the average oscillations of the downward neutrinos}
\item{the energy dependence of the exponential decay term is
in conflict with the data}
\end{itemize}

In the scenario (b), in addition to mixing with $\nu_\tau$,
the unstable component in $\nu_\mu$ decays to some 
sterile state with which it does not mix. 
In this case there is no restriction on \dm and it enters the $\chi^2$ fit as
an independent parameter. We find that 
\begin{itemize}
\item{ The best-fit \dm does not come out naturally to be in the
\dm independent average oscillation  regime 
of $>$ 0.1 $eV^2$, rather  
it is  $0.003 eV^2$.}
\item{The best-fit value of the decay constant $\alpha = 0.3 \times 10^{-5} eV^2$ implying that the decay rate is small so that the mixing angle is 
maximal ($\sin^2 \theta = 0.5$).} 
\item{Large values of $\alpha$ giving complete decay of upward neutrinos 
are also allowed with a high C.L. In fact with 535 days data the best-fit
was in this region.} 
\item{
The 
best-fit value of the decay constant $\alpha$ is non-zero 
signifying that a little amount of decay combined with \dm dependent 
oscillations gives a better fit to the data.}
\end{itemize} 
At the best-fit values of the parameters 
there is no oscillation of the downward neutrinos so that the 
$\cos (2 \pi L/\lambda_{osc})$ term goes to 1. The 
decay term also goes to 1 signifying that there is not 
much decay either for the downward neutrinos and 
the survival probability is $\approx$ 1 
without requiring the mixing angle to be low. 
On the other hand for the upward neutrinos there are oscillations as well as  
little amount of decay. The sub-GeV  
upward neutrinos have smaller oscillation wavelength and they are close to the  
average oscillation limit (survival probability $\sim$ 0.5) 
while for the multi-GeV neutrinos the oscillation wavelength is such that 
one has maximum oscillations
and the survival probability is less than 0.5. Thus this scenario
reproduces the correct energy dependence of the suppression -- namely 
sub-GeV is suppressed less as compared to multi-GeV neutrinos. 
The best-fit value of 
$\alpha$ being even smaller now than the scenario (a) the 
decay term $e^{-\alpha L/E}$  does not vary very
sharply with the zenith distance $L$ or the energy $E$ so that
its wrong energy dependence does not spoil the fit. 

The conversion probability of $\numu$ to $\nutau$ is given by
\begin{equation}
P_{\numu\nutau} = \frac{1}{4} \sin^2 2\theta\{1+\exp(-4\pi L/\lambda_d)
-2 \exp(-2\pi L/\lambda_d) \cos(2\pi L/\lambda_{osc})\}
\label{ntb}
\end{equation}
The value of $P_{\numu\nutau}$ integrated over the energy and
the zenith angle, for $\alpha=0.3\times 10^{-5} eV^2$ 
(the best fit for scenario (b)) is 0.33
for sub-GeV and 0.26 for multi-GeV. 
For $\alpha=0.44\times 10^{-3} eV^2$ (the second minima in the
$\Delta \chi^2$ vs. $\alpha$ curve) 
the corresponding numbers are 0.21 and 0.15, while
for the only $\numu-\nutau$ oscillation case, the
corresponding values are 0.37 and 0.26 respectively. The value of \dm and
\st for both the cases is $0.003 eV^2$ and 1.0 respectively.

The fig. 9 shows that the zenith angle dependence of the scenario (b) is
almost similar to the case of $\nu_\mu - \nu_\tau$ oscillation. 
But the two cases are very different in principle.
For the oscillation case a larger 
$\theta$ implies a larger conversion whereas in scenario (b)
a larger $\theta$ means the fraction of the unstable component is less
in $\nu_\mu$ and the depletion is less. 
If one compares the conversion probability as given by eq.(\ref{ntb}) 
with the one for the $\nu_\mu - \nu_\tau$ oscillation case, then the 
scenario (b)
considered in this paper would have smaller number of $\nu_\tau$s
in the resultant flux at the detector, especially for the larger values 
of $\alpha$ which are still allowed by the data and the two cases might be 
distinguished when 
one has enough statistics to detect $\tau$ appearance in Super-Kamiokande 
\cite{hall} or from neutral current events \cite{smirnov}.

In our paper we have followed the procedure of data fitting as done in
\cite{yasuda}. Thus we use the ratios for which the common systematic
errors get canceled out. 
Strictly speaking one should use the absolute number of events and
include all the correlations between bins and $e$-like and $\mu$-like
events. But the best-fit points and the allowed regions are not
expected to change significantly \cite{private}. 
We have compared the scenarios of neutrino oscillation and
decay with the same definition of $\chi^2$ and for this purpose of
comparison neglecting the correlation matrix will not make much
difference.
Apart from the statistical analysis we have given plots of $R$ and $Y$
for various values of the parameters. The allowed parameter ranges
from these plots are consistent with what we get from our statistical
analysis. The histograms that we have plotted are 
also independent of our definition of $\chi^2$.  We have checked that if  
we estimate the allowed ranges from the histograms these are 
consistent with what we get from our definition of $\chi^2$.
Thus we agree with the observation in ref. \cite{yasuda} that although
this method of data fitting is approximate it works well. 
In our analysis we have used only the SK data because it has the
highest statistics as compared to the earlier atmospheric neutrino
experiments.  

Radiative decays of neutrinos are severely constrained \cite{fukugita}
and what we consider here are the non-radiative decay modes of the neutrino. 
Models for neutrino decay for the scenario (a) are discussed
in \cite{pak,anjan}.  
For the scenario (b) the unstable state decays to a sterile neutrino state and 
a light scalar. 
The model described in \cite{acker} in connection to the solar neutrino problem 
can be adapted to emulate this scenario.
A recent paper \cite{pakvasa2}
has discussed how such a model can be constructed. 
Since in scenario (b) the decay products are invisible
there are no distinctive signs of the decay. 
Decay of leptons to the light scalar are prohibited from conservation of
lepton number. Hence it is 
difficult to constrain these from laboratory experiments \cite{acker}. 
Consequences of such a model for astrophysics has been discussed in
\cite{pakvasa2}. 

\vspace{1cm}

We would like to thank Anjan Joshipura for creating our interest 
in the neutrino decay solution to the atmospheric neutrino problem. 
We also like to thank Osamu Yasuda for many useful correspondences
during the development of our computer code,
S. Pakvasa for useful correspondences  and Kamales
Kar for discussion and encouragement. 
Finally we would like to thank Kate Scholberg for providing us with the
848 days data and for other useful correspondences.

\newpage

\begin{center}
{\bf Figure Captions}
\end{center}

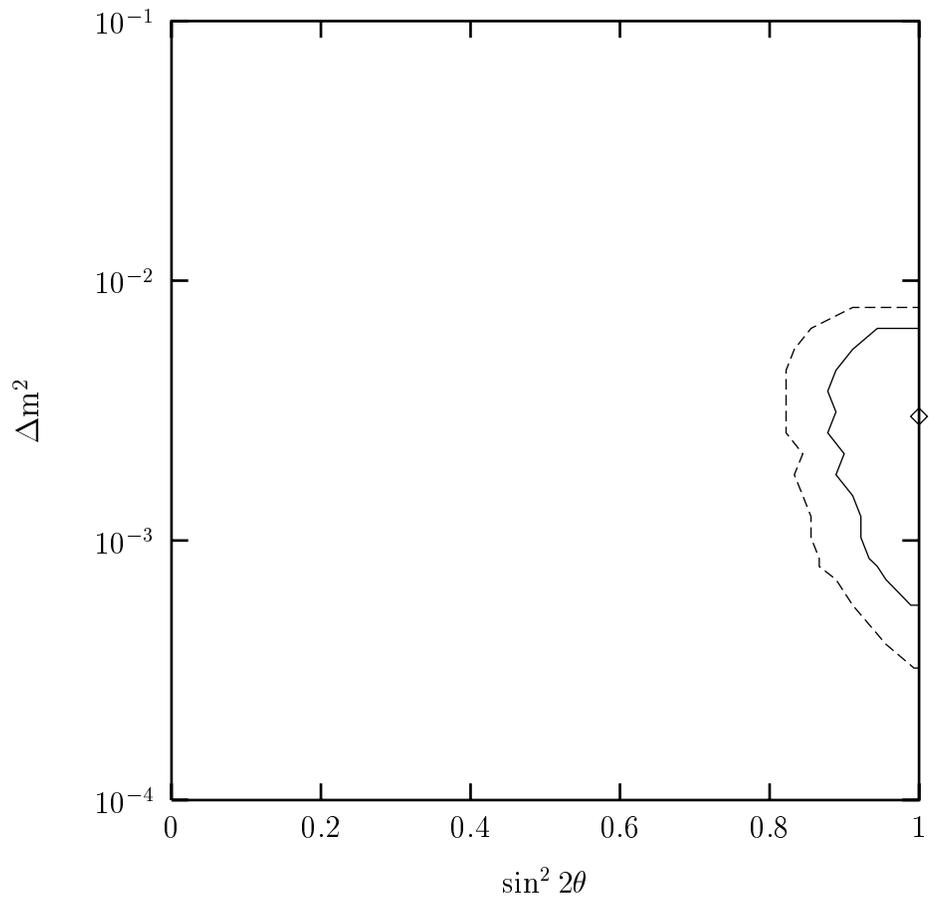
\noindent Fig. 1. The allowed parameter region in \dm-\st plane for the 
$\numu-\nutau$ oscillation hypothesis for the 848 days data. 
The solid line is the area 
allowed at 90\% C.L. and the dashed line shows the area allowed at 
99\% C.L. The best-fit point is shown.

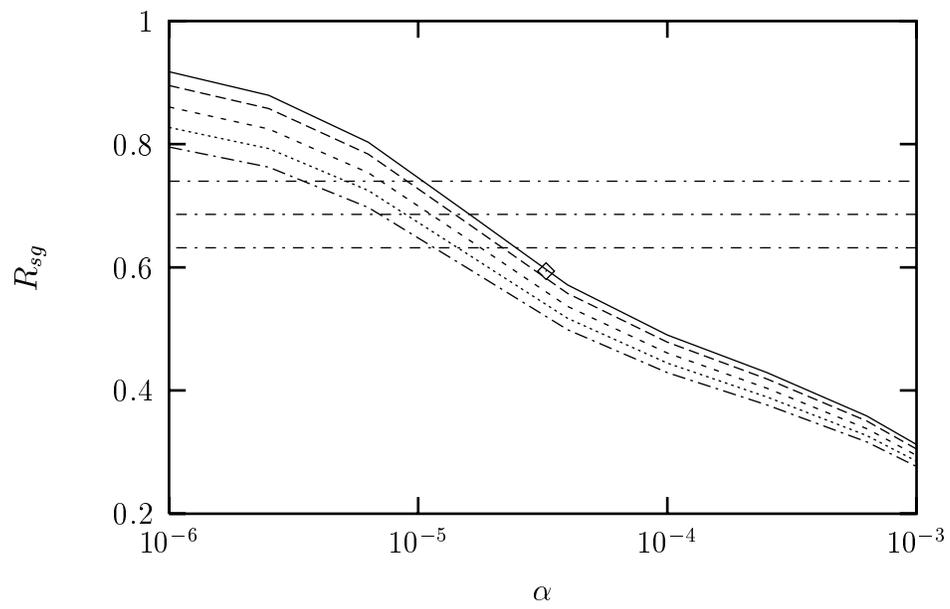
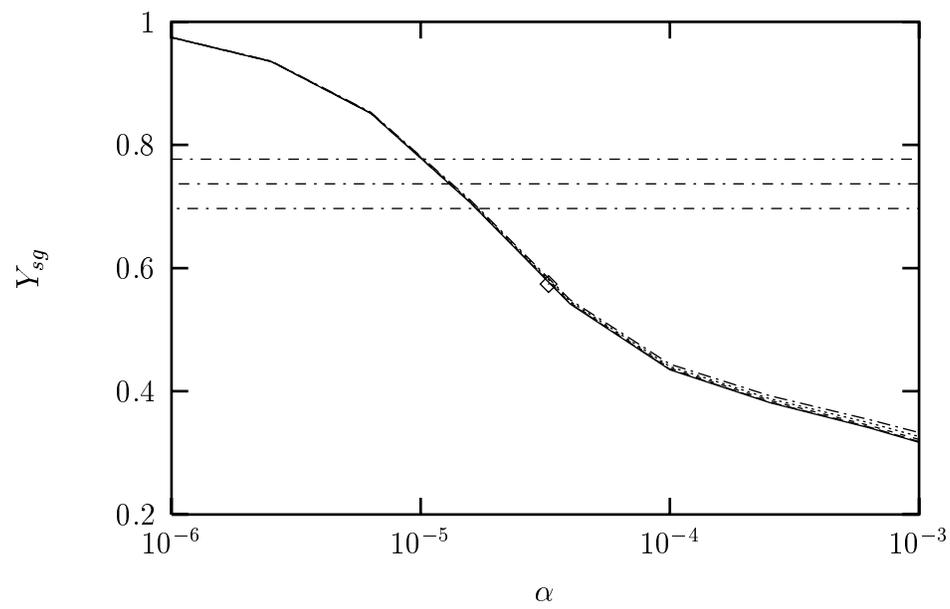
\noindent Fig. 2. The variation of R and Y with $\alpha$ for the sub-GeV 
neutrinos (denoted by the subscript sg) assuming neutrino decay 
with \dm $> 0.1 eV^2$. The curves 
are drawn at fixed values of $\sin^2\theta$=0.03 (solid line), 
$\sin^2\theta$=0.04 (long dashed line), $\sin^2\theta$=0.06 
(short dashed line), $\sin^2\theta$=0.08 (dotted line) 
and $\sin^2\theta$=0.1 (long dashed-dotted line). 
The short dashed-dotted lines give the SK 848 days results within a 
$\pm 1\sigma$ band. Also shown are the R and 
Y at the best-fit point.
  
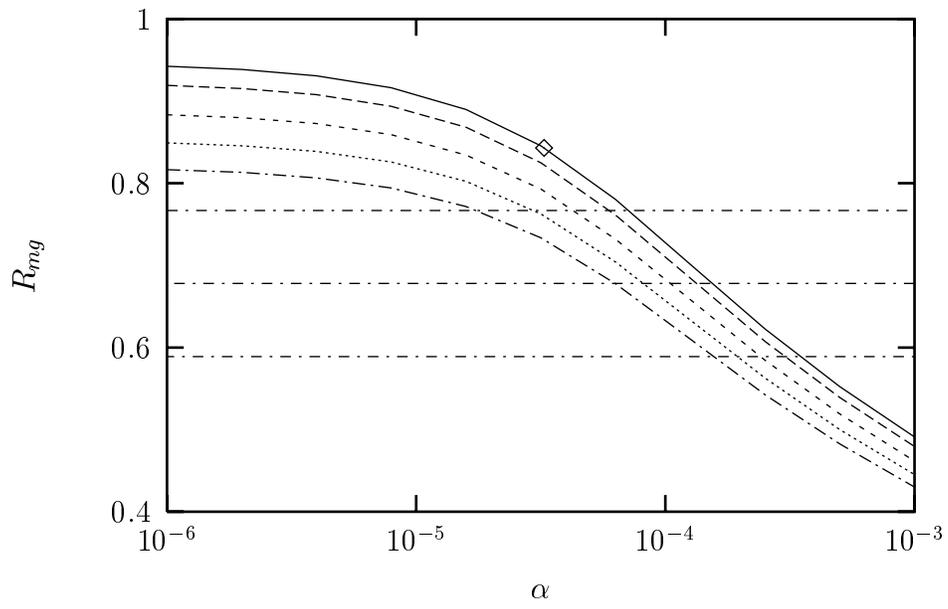
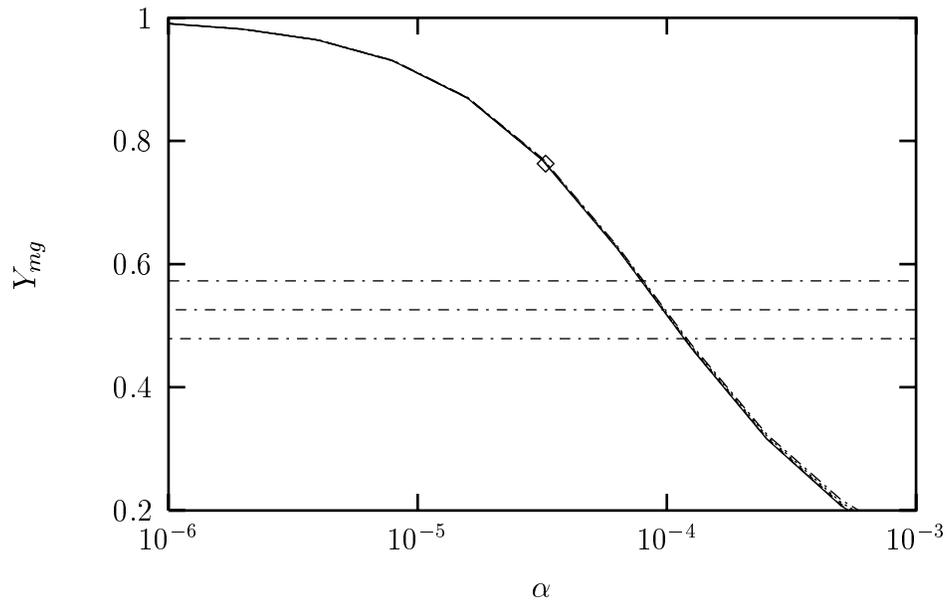
\noindent Fig. 3. Same as in fig. 2 but for multi-GeV neutrinos.
  
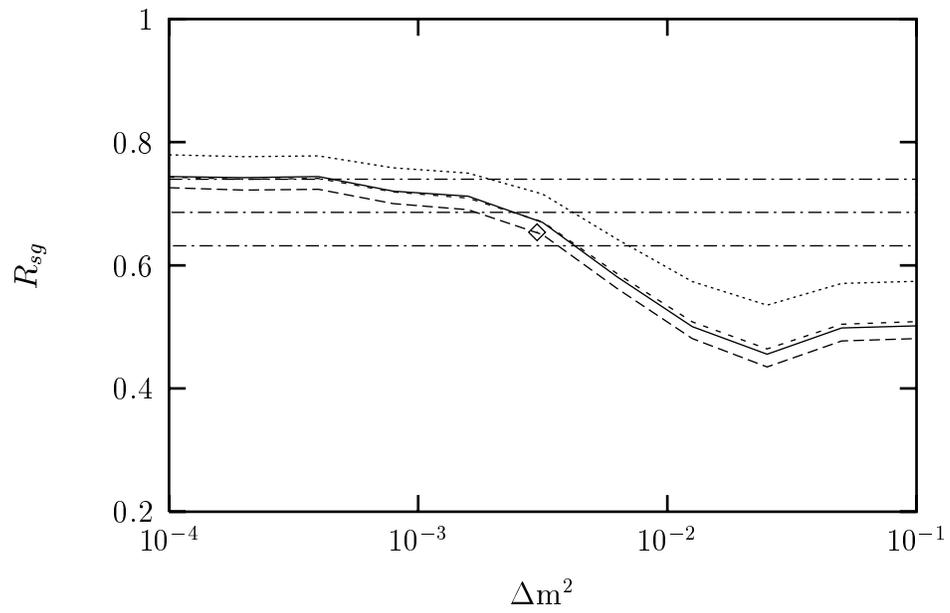
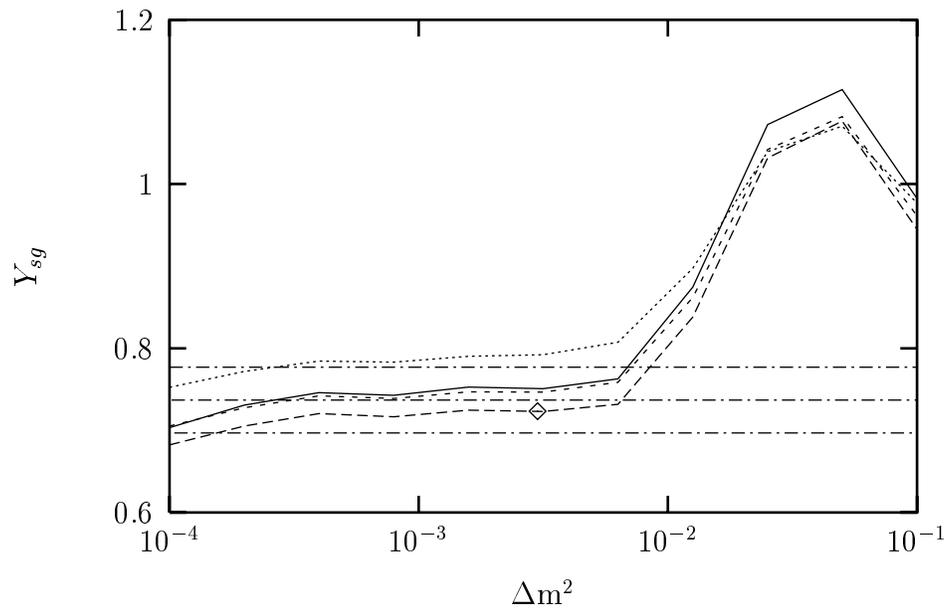
\noindent Fig. 4. The variation of R and Y with \dm for the sub-GeV 
neutrinos (denoted by the subscript sg) 
assuming neutrino decay with \dm unconstrained. In these curves 
the $\alpha$ is fixed at it's best-fit value of $0.3\times 10^{-5} eV^2$.
The curves are drawn at fixed values of $\sin^2\theta$=0.7 (dotted line), 
$\sin^2\theta$=0.6 (short dashed line) and $\sin^2\theta$=0.5 (long 
dashed line). The solid lines give the curves for the best-fit value 
($\sin^2\theta = 0.5$) 
of the $\numu-\nutau$  oscillation case. The dotted-dashed lines give the
SK 848 days results within a $\pm 1\sigma$ band. Also shown are 
the R and Y at the best-fit point.
  
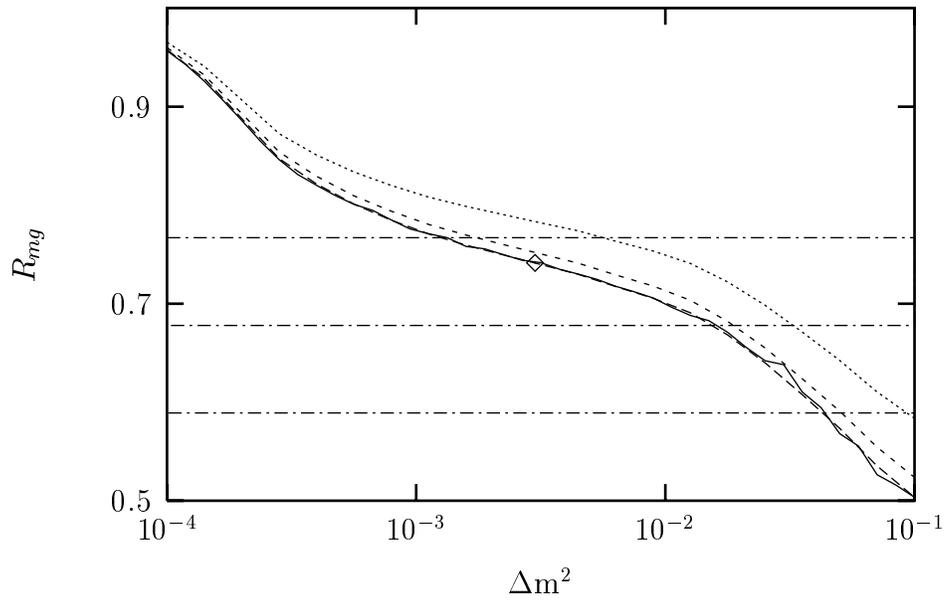
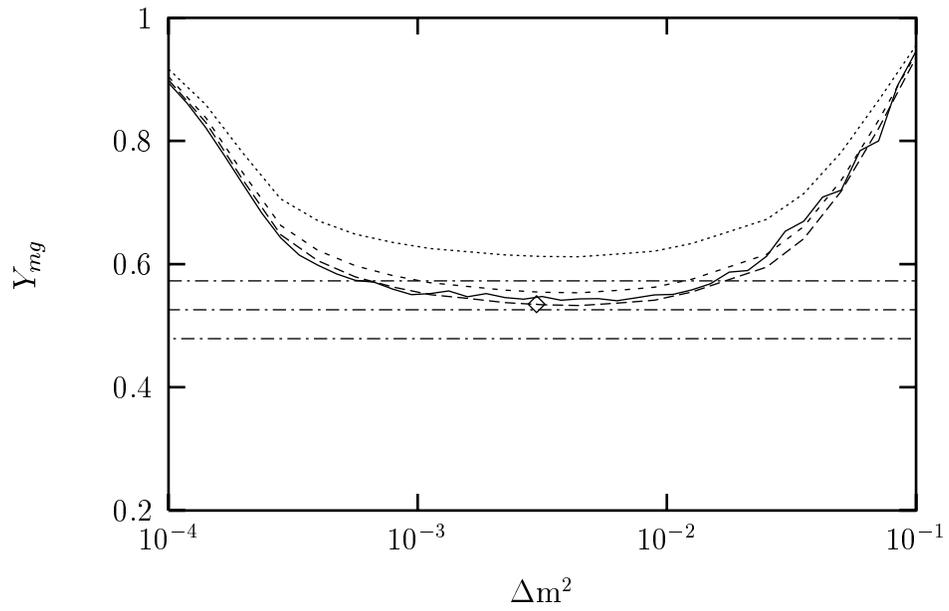
\noindent Fig. 5. Same as in fig. 4 but for multi-GeV neutrinos.
  
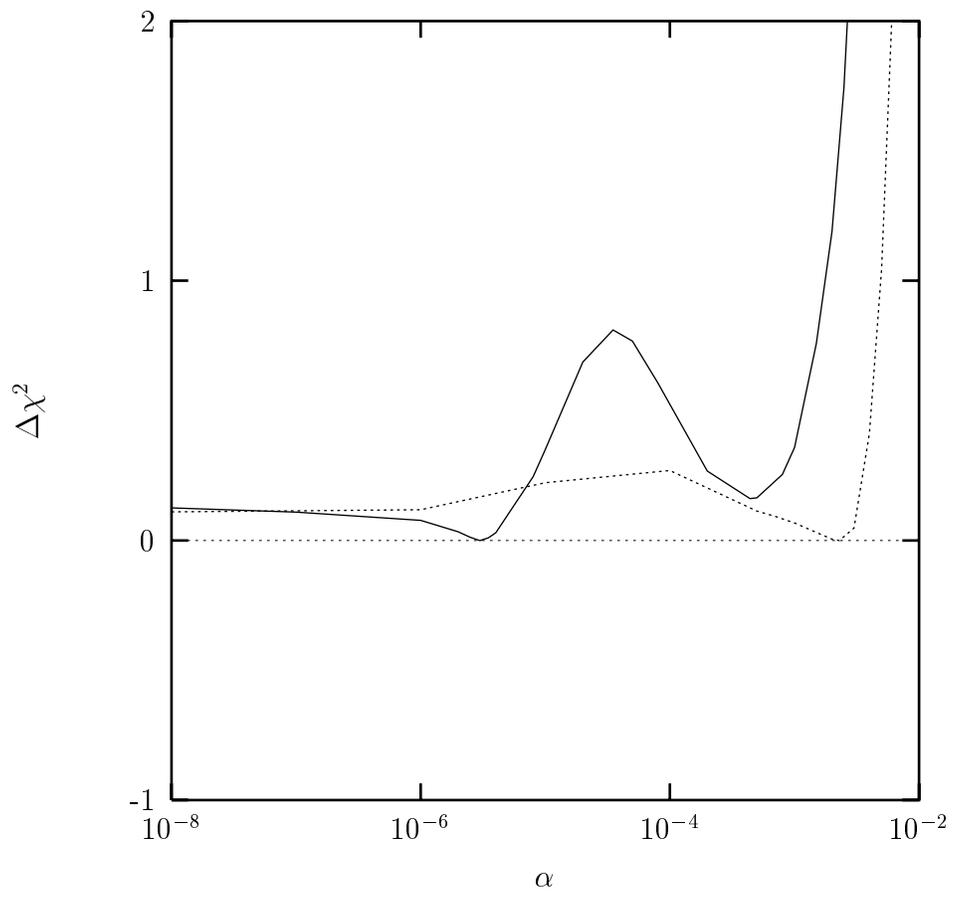
\noindent Fig. 6. The $\Delta \chi^2 = \chi^2 - \chi^2_{\min}$ vs. $\alpha$ 
with \dm and \st unconstrained for the 848 days (solid line) and 
535 days data (dotted line). 

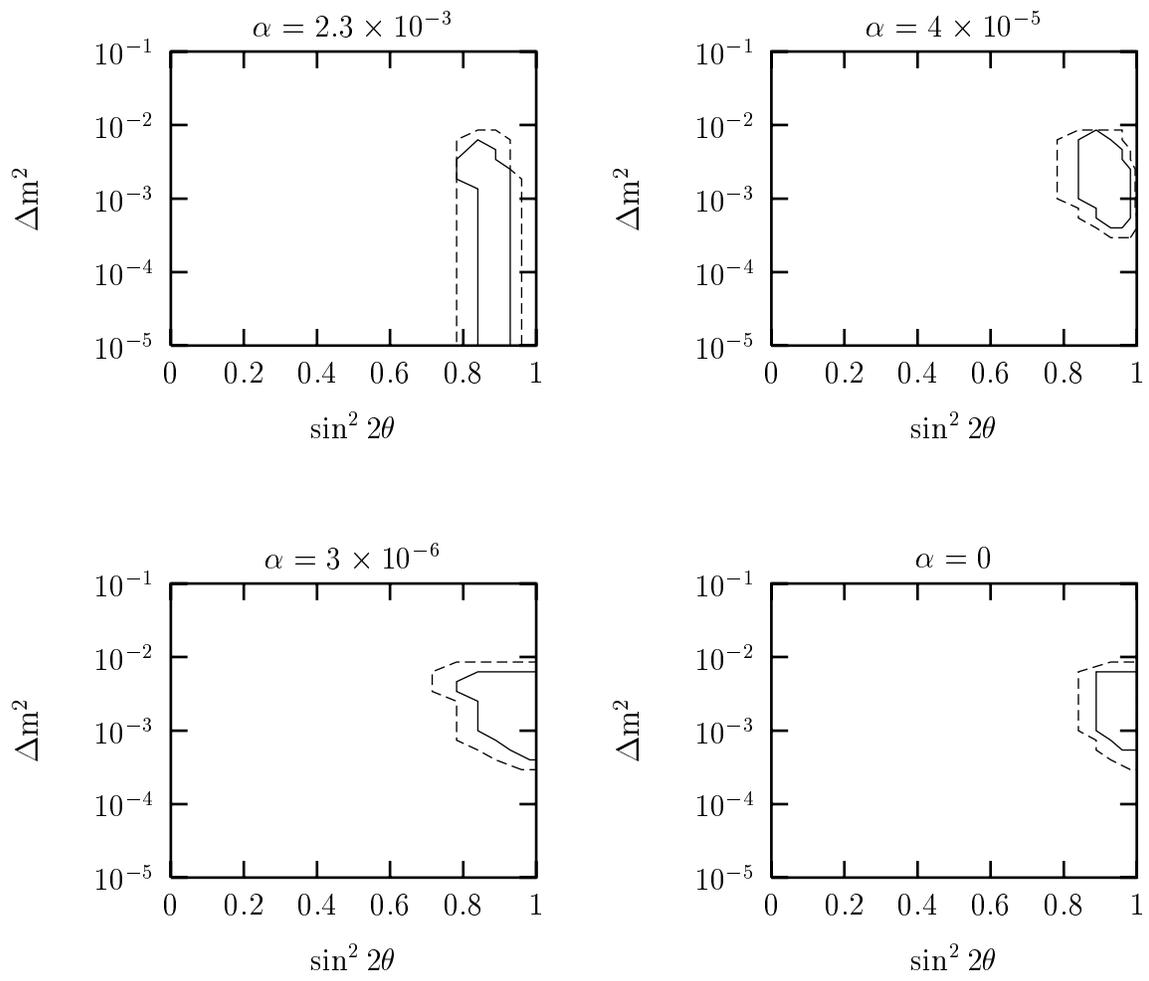
\noindent Fig. 7. The allowed parameter region for the 848 days data 
in the \dm-\st plane for 4 
different values of $\alpha$ shown at the top of each panel. The 
solid and the dashed lines correspond to the area allowed at 90\% C.L. 
and 99\% C.L. respectively.
  
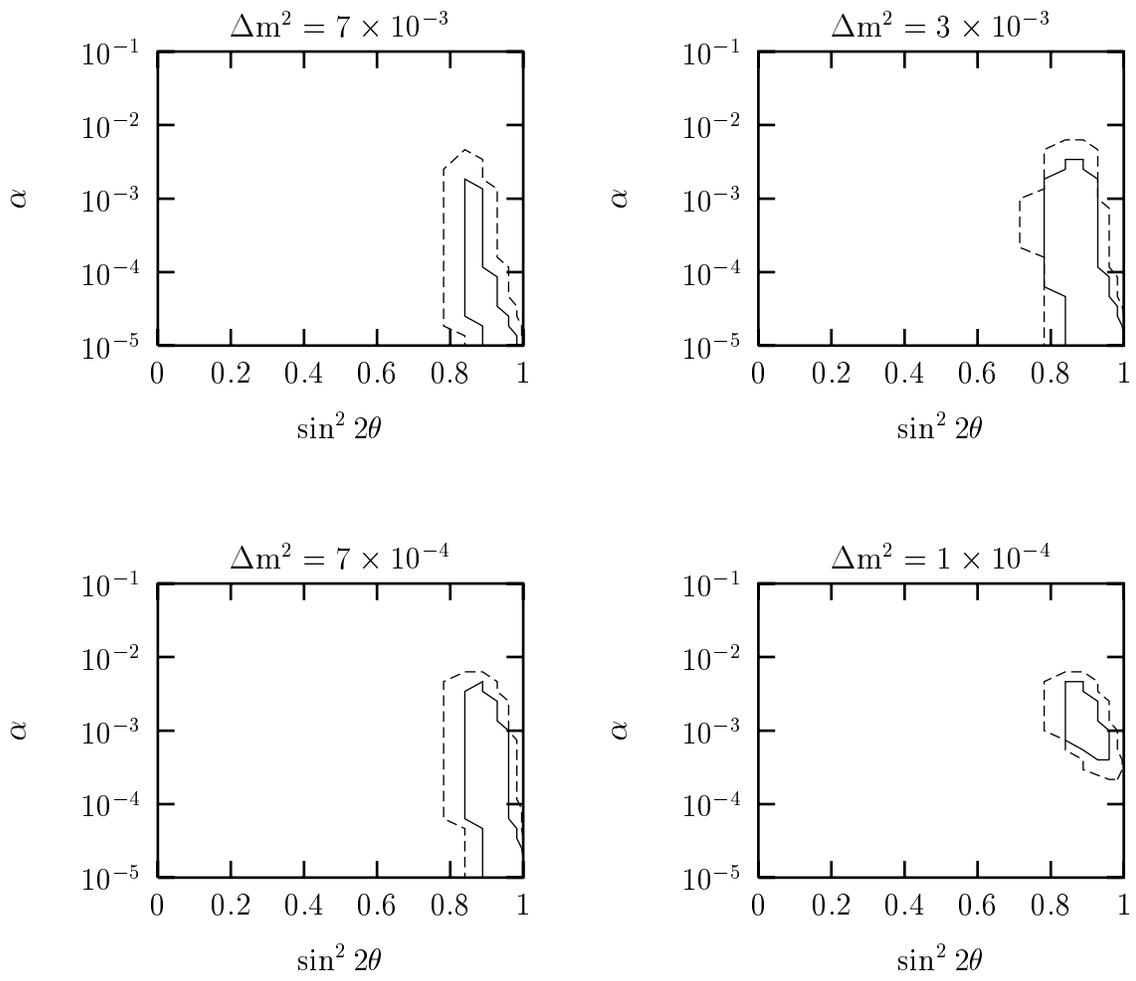
\noindent Fig. 8. The allowed parameter region for the 848 days data 
in the $\alpha$-\st plane 
for 4 different values of \dm shown at the top of each panel. 
The solid and the dashed lines correspond to the area allowed at 
90\% C.L. and 99\% C.L. respectively.
  
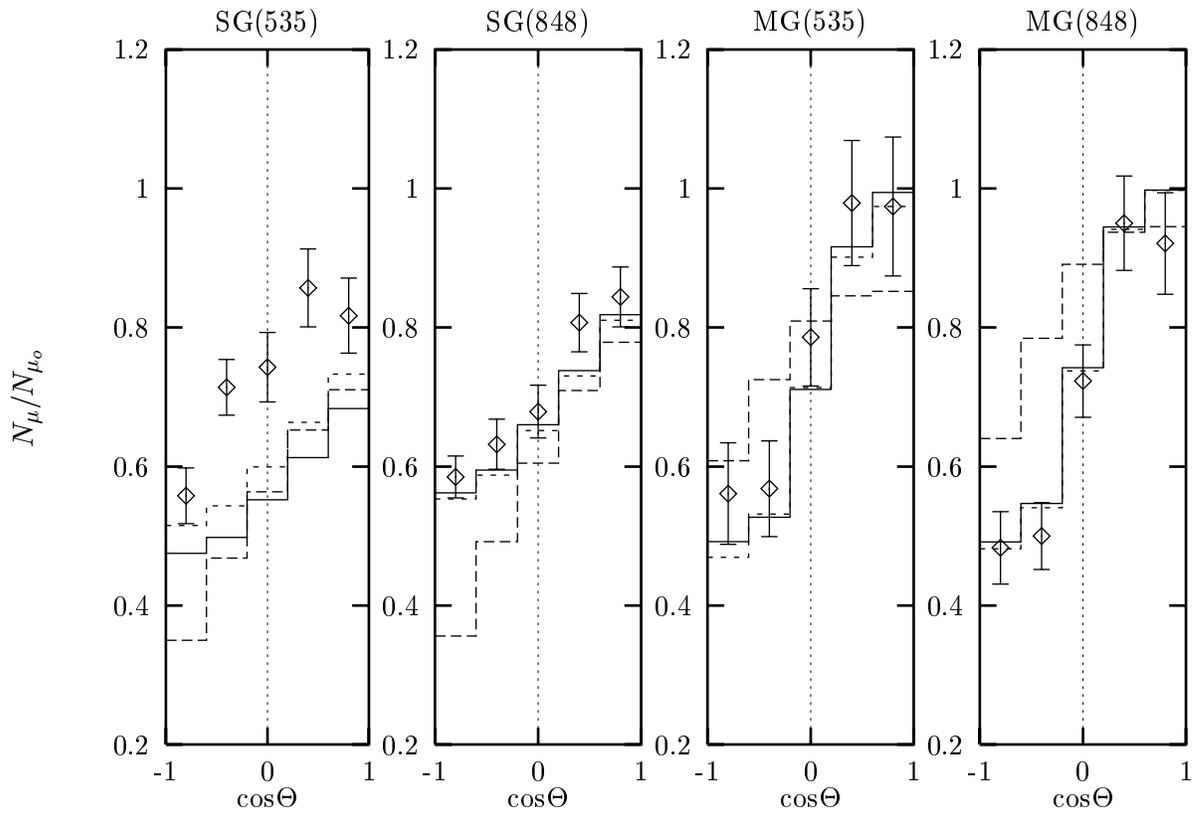
\noindent Fig. 9. The sub-GeV and multi-GeV $\mu$ event distributions vs. 
zenith angle for the various scenarios 
considered. $N_\mu$ is the number of $\mu$ events as given by eq. (3) 
and $N_{\mu 0}$ is the corresponding number with survival probability 1.
The panels labelled SG(535) and MG(535) give the histograms for the sub-GeV 
and multi-GeV 535 days data respectively, 
while the SG(848) and MG(848) give the corresponding 
histograms for the 848 days data. For the both the sets  
the solid line corresponds to the best-fit $\numu-\nutau$ 
oscillation solution, the long dashed line is for the best-fit 
for scenario (a) and the short dashed 
line for the best-fit 
for scenario (b). 
Also shown are the SK $\mu$ event 
distributions with $\pm 1\sigma$ error bars for both the sets.

\end{document}